\documentclass{aa}

\usepackage{graphicx}
\usepackage{natbib}
\bibpunct{(}{)}{;}{a}{}{,}

\title{
Mid-infrared luminosity as an indicator of the total infrared luminosity of 
galaxies
}

\author{
  T.~T.~Takeuchi\thanks{
  Postdoctoral Fellow of the Japan Society for the Promotion of Science for 
  Research Abroad.
}
  \and 
  V.~Buat
  \and
  J.~Iglesias-P\'{a}ramo
  \and
  A.~Boselli
  \and
  D.~Burgarella
}

\institute{Laboratoire d'Astrophysique de Marseille, 
  Traverse du Siphon BP8, 13376 Marseille Cedex 12, France\\
  \email{[tsutomu.takeuchi,veronique.buat,jorge.iglesias,alessandro.boselli,denis.brugarella]@oamp.fr}
}

\offprints{T. T. Takeuchi\\
\email{tsutomu.takeuchi@oamp.fr}.
}

\date{Received/Accepted}

\def\sbs{\object{SBS~0335$-$052}}
\def\arp{\object{Arp~220}}
\def\irasf{\object{IRAS~F10214$+$4724}}
\def\iizw{\object{II~Zw~40}}
\def\lfir{L_{\rm FIR}}
\def\ltir{L_{\rm TIR}}
\def\lrir{L_{\rm TIR2}}
\def\lir{L_{\rm IR}}
\def\lirt{L_{\rm IR}^{\rm total}}
\newcommand{\expc}[1]{\mathsf{E}\left[#1\right]}
\newcommand{\var}[1]{\mathsf{V}\left[#1\right]}

\titlerunning{MIR luminosity of galaxies}
\authorrunning{T.\ T.\ Takeuchi et al.}

\begin{document}

\abstract{
The infrared (IR) emission plays a crucial role in
understanding the star formation in galaxies hidden by dust.
We first examined four estimators of the IR luminosity of galaxies,
$\lfir$ \citep{helou88}, 
$\ltir$ \citep{dale01a}, 
revised version of $\ltir$ \citep{dale02} (we denote $\lrir$),
and $\lir$ \citep{sanders96} by using the observed SEDs of well-known 
galaxies.
We found that $\lir$ provides excellent estimates of the total IR luminosity
for a variety of galaxy SEDs.
The performance of $\lrir$ was also found to be very good.
Using $\lir$, we then statistically analyzed the {\sl IRAS} PSC$z$ galaxy 
sample \citep{saunders00} and found useful formulae relating the MIR 
monochromatic luminosities [$L(12\,\mu\mbox{m})$ and $L(25\,\mu\mbox{m})$] 
and $\lir$.
For this purpose we constructed a subsample of 1420 galaxies with all 
four {\sl IRAS} band (12, 25, 60, and $100\,\mu$m) flux densities.
We found linear relations between $\lir$ and MIR luminosities, 
$L(12\,\mu\mbox{m})$ and $L(25\,\mu\mbox{m})$.
The prediction error with a 95~\% confidence level is a factor of 4--5.
Hence, these formulae are useful for the estimation of the total
IR luminosity only from $12\,\mu$m or $25\,\mu$m observations.
We further tried to make an `interpolation' formula for galaxies at $0<z<1$.
For this purpose we construct the formula of the relation between 15-$\mu$m
luminosity and the total IR luminosity.
We conclude that the 15-$\mu$m formula can be used as an estimator of the
total IR luminosity from $24\,\mu$m observation of galaxies at $z \simeq 0.6$.
}

\maketitle

\section{Introduction}\label{sec:introduction}

Star formation activity is one of the fundamental properties useful to explore
the evolution of galaxies in the universe.
Generally, the star formation rate is measured by the emission from young 
stars, i.e., ultraviolet (UV) and related nebular line emissions.
However, a significant fraction of UV photons are absorbed and re-emitted by 
dust mainly in the infrared (IR), hence the IR emission plays a crucial role
for an understanding of the obscured star formation in galaxies
\citep[e.g.,][]{buat99,buat02,hirashita03}.

Further, clarifying the correlation between flux densities at various
IR bands is an important task to understand the origin, 
release and transfer of energy in galaxies.
Such studies play a crucial role in constructing and verifying IR
galaxy evolution models
\citep[e.g., ][]{granato00,franceschini01,takeuchi01a,takeuchi01b,takagi03}.

Based on their 12-$\mu$m sample of galaxies, \citet{spinoglio95} made a
pioneering study to examine various correlations between flux densities from 
near-IR (NIR) to far-IR (FIR), and presented useful diagnostics for 
Seyferts and normal galaxies on color-color diagrams.
They also found that the 12-$\mu$m luminosity correlates well with 
the bolometric ($0.4\mbox{--}300\,\mu$m) luminosity.

Now that data obtained by {\sl Spitzer} have started to become available, 
we are better able to explore the IR properties of galaxies at high 
redshift.\footnote{URL: {\tt http://www.spitzer.caltech.edu/}.}
The 24-$\mu$m band of {\sl Spitzer} MIPS is very sensitive
\citep[e.g.,][]{papovich04},
and will be used extensively for the studies of high-$z$ galaxies.
Hence, from a practical point of view, it is worthwhile to find a
good estimation method of the total IR luminosity of galaxies from 
the mid-IR (MIR) luminosity.
This will also be useful for forthcoming IR space missions, e.g., ASTRO-F.
\footnote{URL: {\tt http://www.ir.isas.ac.jp/ASTRO-F/index-e.html}.}

In this work, we present the estimation formulae for the FIR luminosity from
the MIR.
We focus on the relation between MIR and total IR luminosities, in contrast to
\citet{spinoglio95}, who used the bolometric luminosity integrated from 
the optical to the IR.
For this purpose, we have to rely on some conventional formulae to estimate 
the total IR luminosity, since direct measurement of the total IR luminosity 
is possible only for a limited number of galaxies.
First we examine the performance of four formulae in use, 
using galaxies with well-measured spectral energy distributions (SEDs).
This sample consists of 17 galaxies ranging from dwarfs to ultraluminous,
and from cool (submillimetre bright) to hot (MIR bright) ones.

We then perform a correlation analysis for the galaxy sample extracted from 
{\sl IRAS} PSC$z$, and obtain a statistical formula for the estimation
of the total IR luminosity from MIR luminosities.
This statistical sample is selected by the criterion that the galaxy has
all four {\sl IRAS} flux density values.
By combining the formula and ISOCAM 15-$\mu$m data, we then give 
an interpolation formula of the FIR luminosity for galaxies at 
$z \simeq 0.6$ observed in the {\sl Spitzer} MIPS 24-$\mu$m band.

The paper is organized as follows:
we examine the four estimators of the total IR luminosity in 
Sect.~\ref{sec:estimator}.
We present our statistical analysis based on {\sl IRAS} PSC$z$ galaxies in 
Sect.~\ref{sec:analysis}.
A reexamination of the estimator $\lir$ and application to galaxies at 
$z \simeq 0.6$ are given in Sect.~\ref{sec:discussion}.
Sect.~\ref{sec:conclusion} is devoted to our conclusions.
The SEDs of observed galaxies used in Sect.~\ref{sec:estimator}
are shown in Appendix~A.
Mathematical details of the regression analysis are presented in 
Appendix~B.

We denote the flux densities at a wavelength $\lambda$ by 
a symbol $S_\lambda$, but the unit is [Jy].
Throughout this work, we assume a flat lambda-dominated low-density universe
with cosmological parameter set $(h,\Omega_0,\lambda_0)=(0.7,0.3,0.7)$,
where $h=H_0/100 \;[\mbox{km\,s}^{-1}\mbox{Mpc}^{-1}]$.

\section{Performance of the estimators for the total IR luminosity}
\label{sec:estimator}

\begin{table}
\centering
 \caption{Well-known galaxy sample.}\label{tab:known_gals}
  \begin{tabular}{lc}
   \hline
    Name & References$^{\mathrm{a}}$     \\
   \hline
    \multicolumn{2}{c}{Normal galaxies 
    ($10^{9} L_\odot <{\lirt}^\mathrm{b}<10^{11}L_\odot $)}\\
   \hline
    \object{M 63} & 1 \\
    \object{M 66} & 1 \\
    \object{M 82} & 1 \\
    \object{M 83} & 1 \\
    \object{NGC 891} & 1 \\
    \object{NGC 3079} & 1 \\
    \object{NGC 4418} & 1 \\
    \object{NGC 7714} & 1 \\
   \hline
    \multicolumn{2}{c}{IR luminous galaxies ($\lirt > 10^{11}L_\odot$)}\\
   \hline
    \object{NGC 2623} & 1,2 \\
    \object{NGC 7679} & 1,2 \\
    \object{UGC 2982} & 1,2 \\
    \object{UGC 8387} & 1,2 \\
    \arp & 1,3 \\
    \irasf & 1,4 \\
   \hline
    \multicolumn{2}{c}{Dwarf galaxies ($\lirt < 10^{9}L_\odot$)}\\
   \hline
    \object{NGC 1569} & 1,5 \\
    \iizw$^{\mathrm{c}}$ & 1,6 \\
    \sbs$^{\mathrm{c}}$ & 1,7,8\\
   \hline
 \end{tabular}
\begin{list}{}{}
\item[$^{\mathrm{a}}$] References: 
   1. NED, 
   2. \citet{dunne01}, 
   3. \citet{downes93},
   4. \citet{downes92},
   5. \citet{galliano03},
   6. \citet{madden00}, and
   7. \citet{dale01b}.
   8. \citet{houck04}.
\item[$^{\mathrm{b}}$] Total IR luminosity $\lirt$ is calculated by 
integrating over the wavelength range of $\lambda = 8\mbox{--}1000\,\mu$m.
\item[$^{\mathrm{c}}$] The longest wavelength flux densities are 
calculated by an extrapolation using the model of \citet{takeuchi03a}.
\end{list}
\end{table}

Since direct measurement of the total IR luminosity is only available for 
a limited number of galaxies, we have to use a formula to estimate 
the total IR luminosity from discrete photometric data, mainly in the 
{\sl IRAS} bands.
In this section, we examine the performance of four formulae in use.

\subsection{Estimators}

First, we define ${\cal L}_\nu$ as the luminosity per unit frequency at
a frequency $\nu = c/\lambda$ ($c$: the speed of light).
The unit of ${\cal L}_\nu$ is $[\mbox{erg\,s}^{-1}\mbox{Hz}^{-1}]$
throughout this work.

We examine the following four total IR luminosity estimators.
\begin{enumerate}
\item The classical FIR luminosity between 
$\lambda = 42\mbox{--}122\,\mu\mbox{m}$ \citep{helou88}, defined as
\begin{eqnarray}
  \lfir &\equiv& 3.29 \times 10^{-22} \nonumber \\
   &&\times \left( 
   2.58{\cal L}_\nu(60\mu\mbox{m}) + {\cal L}_\nu(100\mu\mbox{m}) \right)
   \;[L_\odot] \;.
\end{eqnarray}
\item The `total' IR luminosity, TIR ($\lambda=3\mbox{--}1100\,\mu$m), 
advocated by \citet{dale01a}
\begin{equation}
  \ltir \equiv \lfir \times 10^{a_0+a_1x+a_2x^2+a_3x^3+a_4x^4} \;[L_\odot]
\end{equation}
where $x\equiv \log (S_{60}/S_{100})$, and $(a_0,a_1,a_2,a_3,a_4)=
(0.2738, -0.0282, 0.7281, 0.6208, 0.9118)$.
\item The updated version of the TIR ($\lambda=3\mbox{--}1100\,\mu$m)
presented by \citet{dale02} (here we denote $\lrir$)
\begin{eqnarray}
  \lrir &\equiv 2.403& \nu{\cal L}_\nu(25\mu\mbox{m})
    -0.2454 \nu{\cal L}_\nu(60\mu\mbox{m}) \nonumber \\
    &&+1.6381 \nu{\cal L}_\nu(100\mu\mbox{m}) \;[L_\odot].
\end{eqnarray}
This is better calibrated for submillimeter wavelengths than $\ltir$.
\item The luminosity between $\lambda = 8\mbox{--}1000\,\mu\mbox{m}$ 
presented by \citet{sanders96}. 
In this work we refer to their IR luminosity estimator as $\lir$.
\begin{eqnarray}
  \lir &\equiv& 4.93 \times 10^{-22} \left[
    13.48{\cal L}_\nu (12\mu\mbox{m})+5.16{\cal L}_\nu (25\mu\mbox{m})
    \right.\nonumber \\
  &&+2.58{\cal L}_\nu (60\mu\mbox{m})+
    \left.{\cal L}_\nu (100\mu\mbox{m})\right] \;[L_\odot] \;.
\end{eqnarray}
\end{enumerate}

\subsection{Examination of the IR luminosity estimators by known galaxies}

\begin{figure}
\resizebox{\hsize}{!}{
\includegraphics[width=0.7\textwidth]{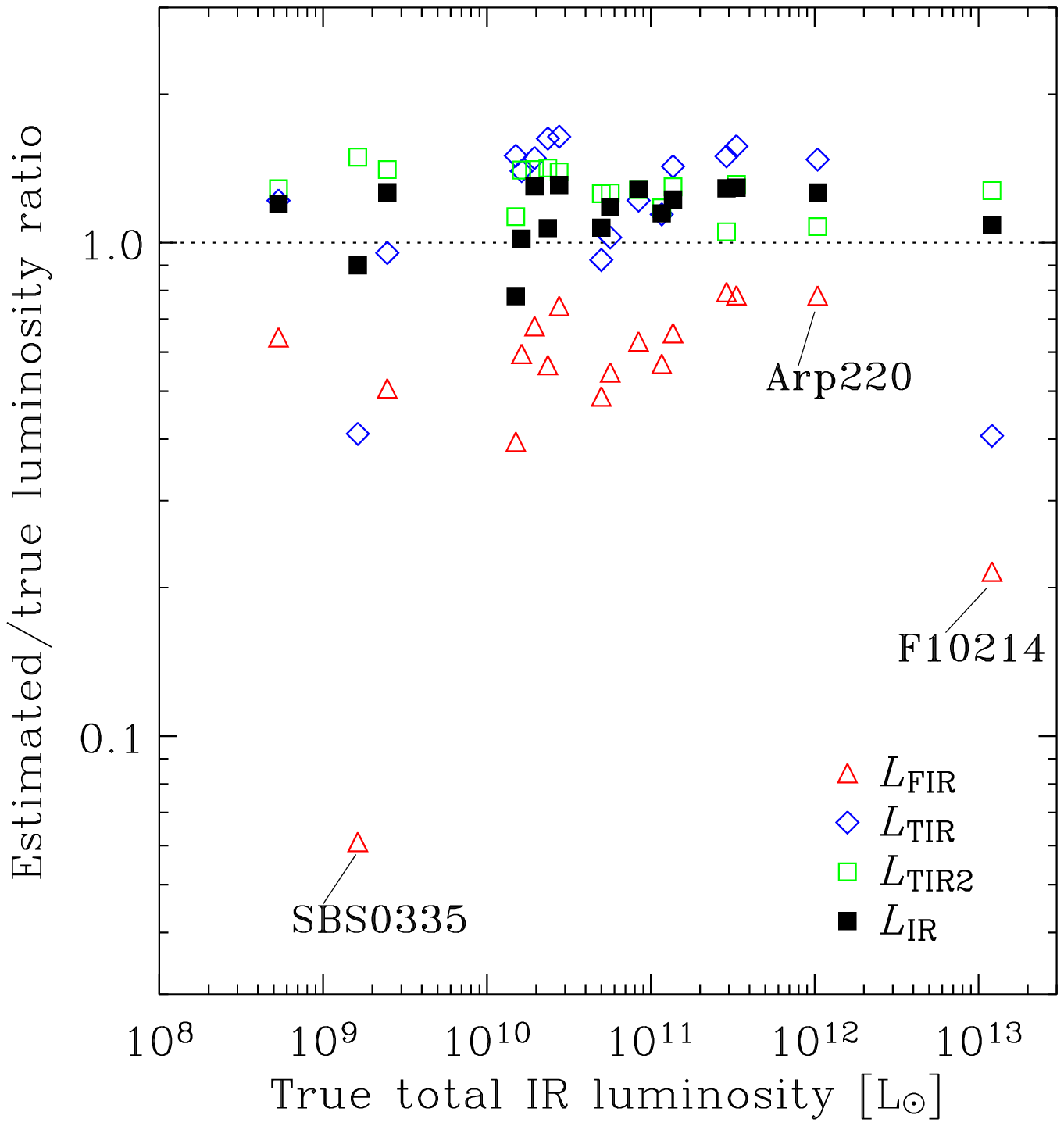}
}
\caption{Comparison between the true total infrared luminosity and
the estimated luminosity of galaxies.
Open triangles, open diamonds, open squares and filled squares represent
the ratios $\lfir/\lirt$, $\ltir/\lirt$, $\lrir/\lirt$, and $\lir/\lirt$,
respectively.
}\label{fig:total}
\end{figure}

Though these IR luminosity estimators are popular in related fields,
direct comparison between the measured IR luminosity and the estimated
value has rarely been done to date.
We examine the performance of the above estimators using
the SEDs of observed galaxies.
We compiled 17 galaxies with well-measured flux densities, with a total
IR luminosity range of $10^{8}L_\odot \la \lirt \la 10^{13}L_\odot$
(see Table~\ref{tab:known_gals}).
Among the dwarf galaxy sample ($\lirt < 10^{9}L_\odot$), the longest 
wavelength data (i.e., FIR and submm) are not available for \iizw\ and \sbs.
We calculated the flux densities by extrapolating their SEDs using
the model of \citet{takeuchi03a} \citep[see also][]{takeuchi04}.
The compiled SEDs are presented in Appendix~A.

We calculated $\lirt$ by integrating the observed data directly within a
wavelength range of $\lambda = 8\mbox{--}1000\,\mu$m by interpolation and 
extrapolation.
Figure~\ref{fig:total} shows the comparison between $\lirt$ 
and estimated luminosities of galaxies.
As expected, the classical $\lfir$ gives systematically lower luminosities
than the true ones, because it represents the luminosity at $42\mbox{--}
122\,\mu$m, and therefore the MIR and submm radiations are not included.
Especially, two galaxies with hot dust (\sbs\ and \irasf) significantly
deviate downward from the diagonal line.

\citet{dale01a} considered the correction factor for the contribution
outside the range of $\lfir$ as a function of the ratio $S_{60}/S_{100}$.
We see that the estimation is clearly improved, but the IR luminosities of 
the two extreme objects are still underestimated.
This is because their $\ltir$ has been designed for normal 
galaxies, and not for such extreme objects.

In contrast to the above two estimators, 
$\lrir$ and $\lir$ give much better 
estimates for all the galaxies in Table~\ref{tab:known_gals}.
They work not only for the objects with very hot dust emission
like \irasf\ and \sbs, 
but also for a heavily extinguished galaxy like \arp.
For \sbs, $\lir$ gives a better result.
This is an expected result because $\lrir$ uses three 
(25, 60, and $100\,\mu$m), and $\lir$ uses all four {\sl IRAS} flux
densities.
In general, $\lrir$ gives slightly larger values than 
$\lir$ does, probably because the considered wavelength range for the former 
($\lambda = 3 \mbox{--}1100\,\mu$m) is
wider than that for the latter ($\lambda = 8 \mbox{--}1000\,\mu$m).

Thus, $\lir$ is the best estimator of the total IR luminosity.
As long as we have the four {\sl IRAS} flux densities, we can 
obtain a precise estimate for the total IR luminosity.
When data in three (25, 60 and $100\,\mu$m) or two (60 and $100\,\mu$m) 
bands are available, $\lrir$ and $\ltir$ give reasonable values except 
for galaxies with extremely hot dust.
$\lrir$ works almost as accurately as $\lir$.
In the following discussions we regard $\lir$ as the correct estimate of
$\lirt$ and use $\lir$ as $\lirt$ itself.

\section{Statistical analysis of the {\sl IRAS} sample}\label{sec:analysis}

Our next step is to find a conventional formula to estimate $\lir$ only from 
a single MIR band.
For this purpose, we make a regression analysis for $\lir$ and MIR
luminosities in the {\sl IRAS} bands.
Here we define the luminosity at a wavelength $\lambda$, $L(\lambda)$, as
\begin{equation}
  L(\lambda) \equiv \lambda {\cal L}_\lambda = \nu {\cal L}_\nu \,,
\end{equation}
and we discuss $L(12\,\mu\mbox{m})$ and $L(25\,\mu\mbox{m})$.
Mathematical details of the regression analysis can be found in 
Appendix~B.

\subsection{{\sl IRAS} Sample}\label{subsec:sample}

We selected a sample from {\sl IRAS} PSC$z$ 
\citep[hereafter PSC$z$,][]{saunders00}.
The PSC$z$ is a complete, flux-limited all-sky redshift survey catalog of 
{\sl IRAS} galaxies with a detection limit of $S_{60} > 0.6$~Jy.
It contains 15411 {\sl IRAS} galaxies with redshifts.
Out of the whole sample, we selected galaxies with good quality flux 
densities for all four {\sl IRAS} bands (12, 25, 60, and 
$100\,\mu$m) for this analysis, because $\lir$ requires all four 
flux densities.
We performed this procedure as follows:
1.\ We examined the flux origin and quality flags given in PSC$z$ 
for the point source flux density, and omitted galaxies with upper limits 
(denoted as 1 in {\sf pscz.dat}),
2.\ We extracted the coadded or extended addscan flux densities.
We adopted this selection because we found that the addscan/coadded
fluxes with quality flag 1 include unrealistic values close to
the upper limits in point source flux densities.

There is a caveat that the selection by using all the four {\sl IRAS} 
bands would introduce a subtle sample bias in the analysis.
In order to see if the bias is serious, we also made a subsample by omitting
the galaxies with flag 1 only at $100\,\mu$m (3260 galaxies included).
This subsample for comparison gave essentially the same result as
the above sample (the difference was less than $\sim 1\,\%$).
This means that the selection in the MIR affects the result only very slightly,
and the sample properties are controlled by the FIR.
It is a clear contrast to the sample of \citet{spinoglio95} which was 
12-$\mu$m selected: the present sample consists of more quiescent, normal
galaxies than theirs.
A full treatment including the upper-limit sample will be presented 
elsewhere (Takeuchi et al.\ 2004, in preparation).
Our final subsample contains 1420 galaxies.

\subsection{Results}\label{subsec:results}

\begin{figure}
\resizebox{\hsize}{!}{
\includegraphics[width=0.7\textwidth]{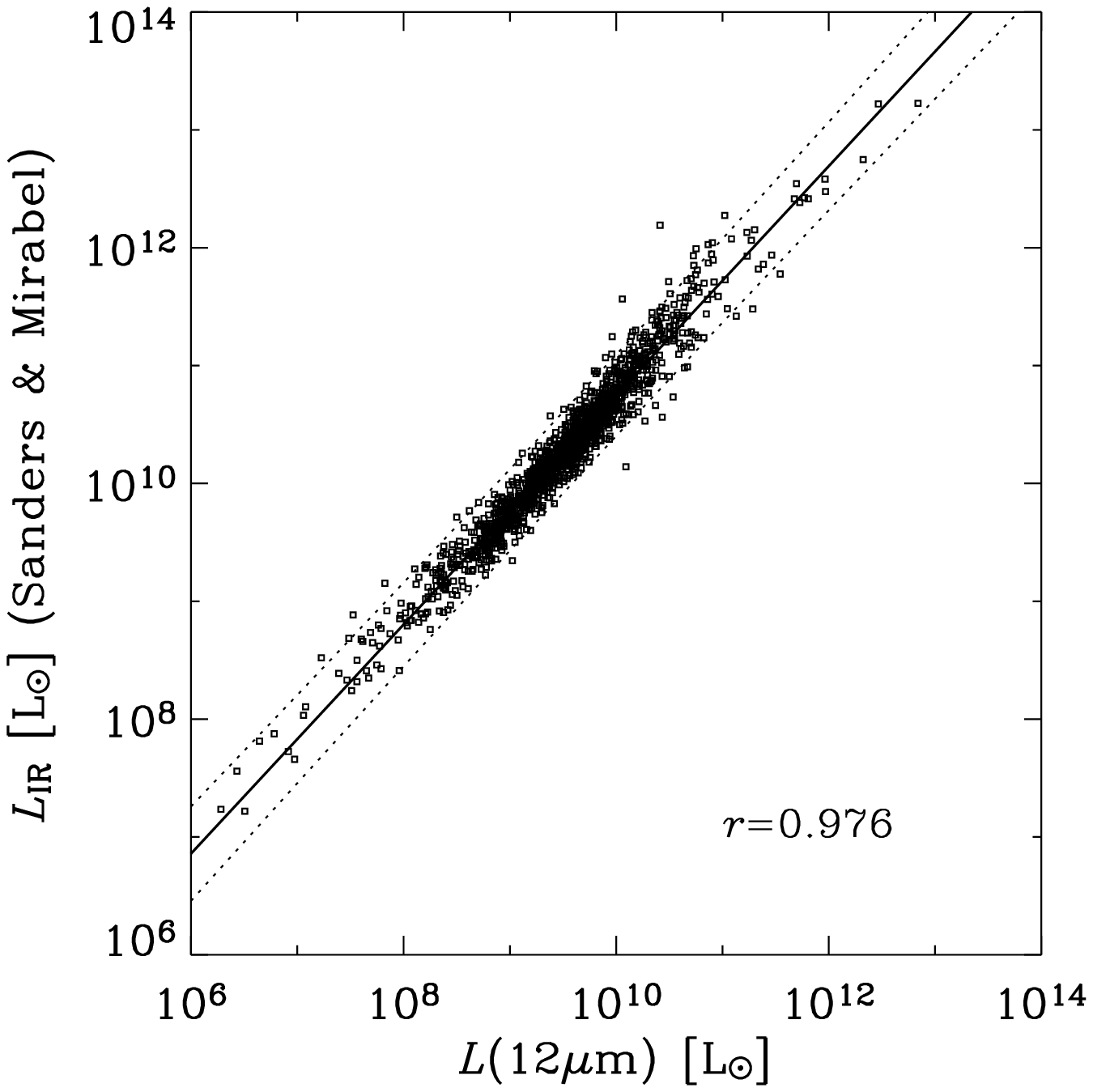}
}
\caption{The relation between $12\,\mu$m luminosity $L(12\,\mu\mbox{m})$ 
and $\lir$, the estimate of the total infrared (IR) luminosity $\lirt$ 
proposed by \citet{sanders96}.
}\label{fig:12um}
\end{figure}

\subsubsection{The $L(12\,\mu\mbox{m})$--$\lir$ relation}

For the $12\,\mu$m luminosity, we obtained the regression parameters 
for $\lir$ as follows:
\begin{eqnarray}
 \log \lir &=& 1.02 + 0.972 \log L(12\,\mu\mbox{m}) \;, \label{eq:12um}\\
 r &=& 0.976 \;, \\
 V_{\rm e} &=& 0.0238\;,
\end{eqnarray}
where $r$ is the correlation coefficient and $V_{\rm e}$ is the dispersion
in the linear model (see Appendix~B).
Here the above $V_{\rm e}$ gives the 95~\% confidence interval 0.3--0.4.
The data points and the regression line are shown in Fig.~\ref{fig:12um}.
The 95~\% confidence limits for the prediction error are presented by 
dotted lines.

We see a tight linear relation between $L(12\,\mu\mbox{m})$ and $\lir$, with a
correlation coefficient $r=0.976$.
As seen in Sect.~\ref{sec:estimator}, the scatter in Fig.~\ref{fig:12um} 
is not due to the estimation error, but is caused by the intrinsic
properties of individual galaxies: it is a reflection of the physical
variety in the SEDs of the sample galaxies.
We will discuss the origin of the scatter in future work
(Takeuchi et al.\ 2004 in preparation).
It gives the prediction error of a factor of 4--5 at the
IR luminosity range $[10^6L_\odot,10^{11}L_\odot]$.
It is an interesting result because we know there is a large variety of IR 
SEDs among galaxies, depending on their activities.

\subsubsection{The $L(25\,\mu\mbox{m})$--$\lir$ relation}

\begin{figure}
\resizebox{\hsize}{!}{
\includegraphics[width=0.7\textwidth]{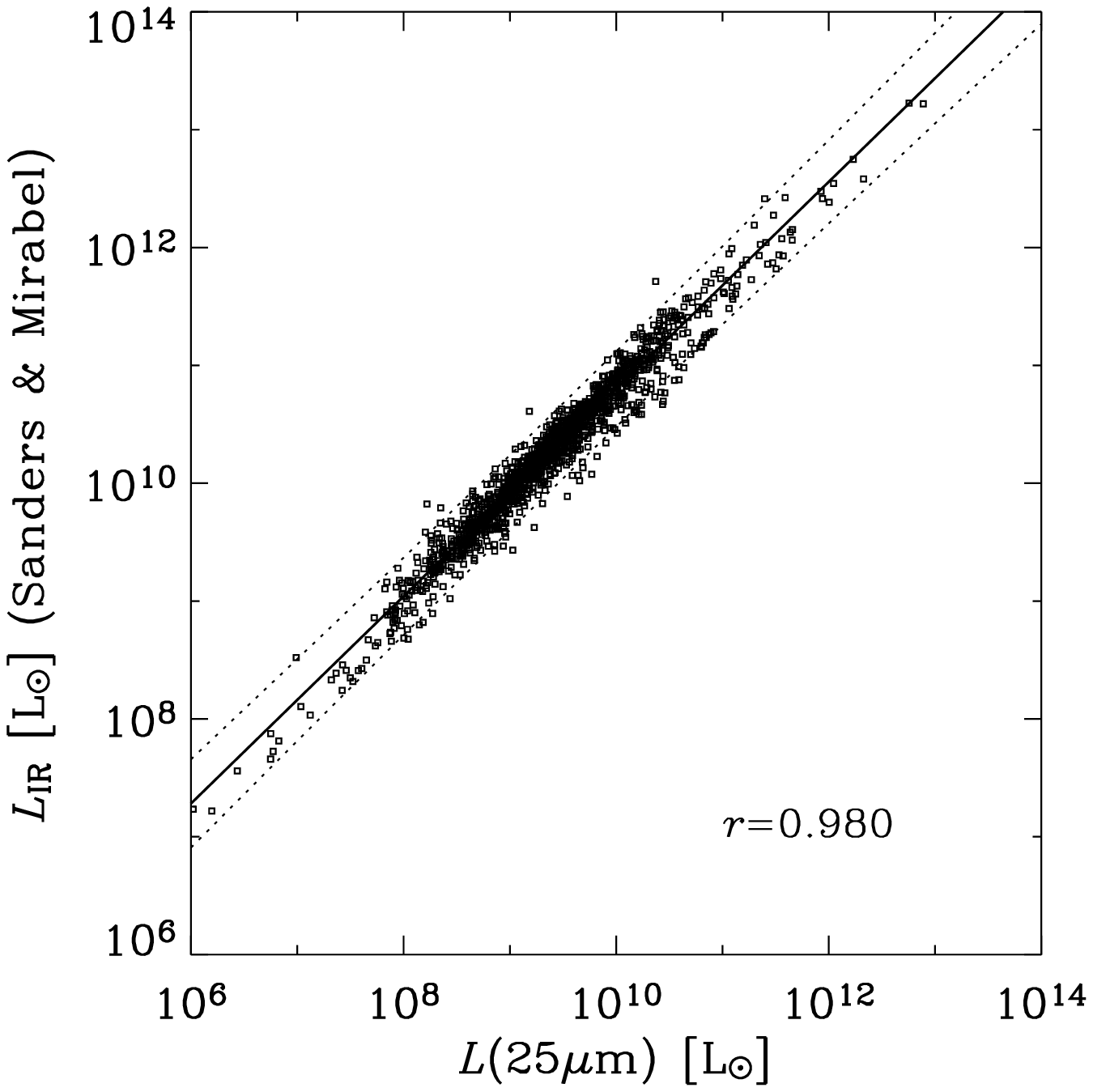}
}
\caption{The relation between $25\,\mu$m luminosity $L(25\,\mu\mbox{m})$ and 
$\lir$.
}\label{fig:25um}
\end{figure}

As above, for the $25\,\mu$m luminosity, we obtained the 
regression parameters for $\lir$ as
\begin{eqnarray}
 \log \lir &=& 2.01 + 0.878 \log L(25\,\mu\mbox{m}) \;, \label{eq:25um}\\
 r &=& 0.980 \;, \\
 V_{\rm e} &=& 0.0203\;.
\end{eqnarray}
This yields the 95~\% confidence interval of 0.3--0.5.
The data points and the regression lines are shown in Fig.~\ref{fig:25um}
Again, the 95~\% confidence limits are presented by dotted lines.
The width of the confidence interval corresponds to a factor of 4--6.

Thus, we conclude that both $L(12\,\mu\mbox{m})$ and $L(25\,\mu\mbox{m})$ 
provide us with reliable estimates for the total IR luminosity $\lir$, 
which are valid for several orders of magnitude in IR luminosity.

\section{Discussion}\label{sec:discussion}

\subsection{Applicability and limitation of the linear relations}
\label{subsec:applicability}

In Sect.~\ref{sec:analysis}, we obtained fairly tight linear relations 
between MIR luminosities $L(12\,\mu\mbox{m})$ and $L(25\,\mu\mbox{m})$, 
and $\lir$.
We also found that the scatter in the relations is due to the intrinsic
properties of the SEDs of galaxies,
and we see some galaxies significantly deviating from the 95~\% 
confidence intervals.
Then, a natural question is: for which type of galaxy does the
relation work well?
Among the sample galaxies in Table~\ref{tab:known_gals}, we have some 
galaxies with SEDs indicative of warm or hot dust
(\sbs, \iizw, and \irasf), as well as those
with SEDs indicative of cold dust (\object{NGC 1569} and \arp).
In order to examine the applicability and limitation of the relations, we
revisit the well-known galaxy sample presented in Table~\ref{tab:known_gals}.
We represent the luminosity predicted from the linear relations 
[Eqs.~(\ref{eq:12um}) and (\ref{eq:25um})] by $\lir^{\rm linear}$.

\begin{figure}
\resizebox{\hsize}{!}{
\includegraphics[width=0.7\textwidth]{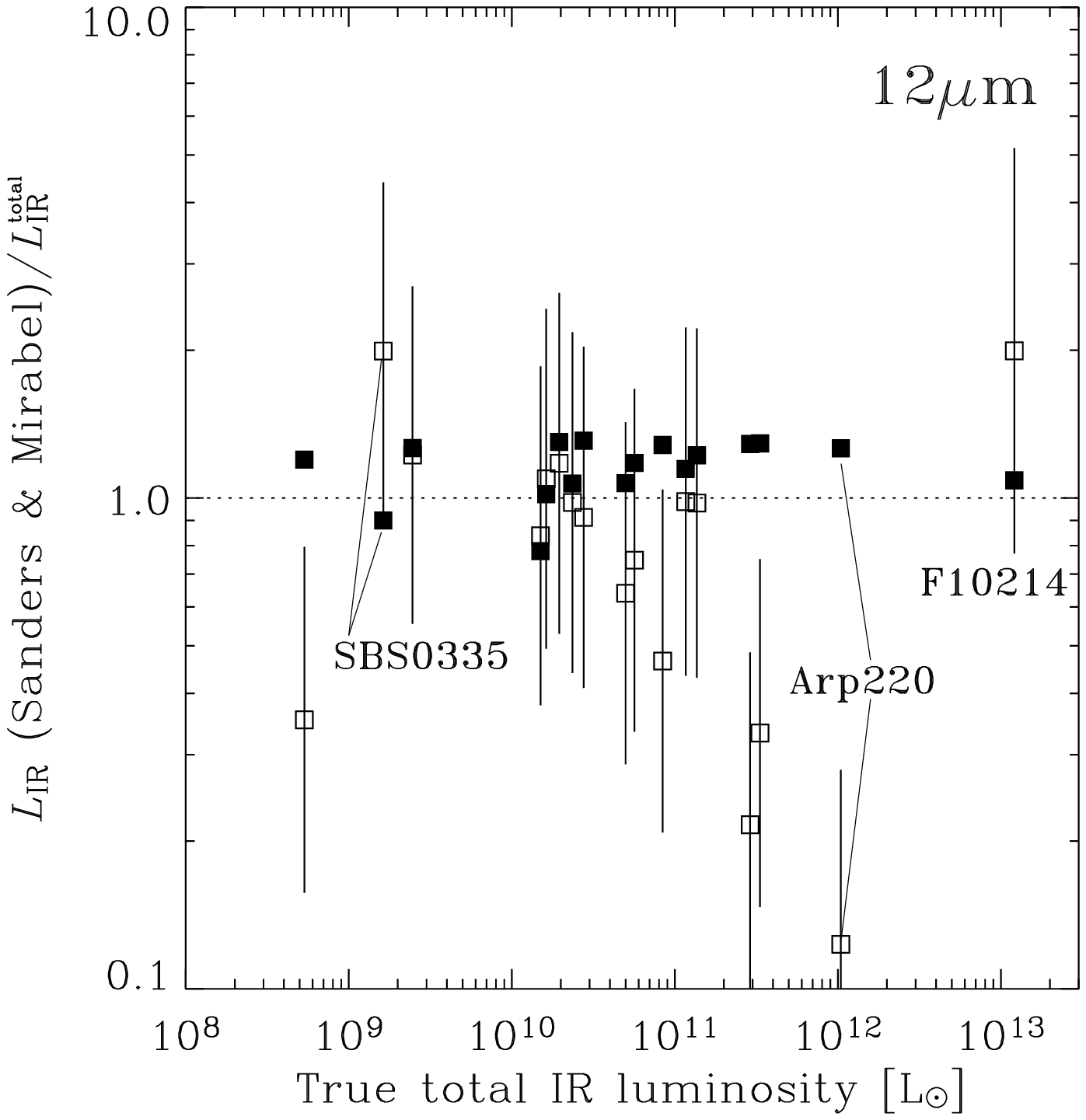}
}
\caption{The performance of the linear estimation formula obtained from the 
$L(12\,\mu\mbox{m})$--$\lir$ relation.
Filled squares are the same as those in Fig.~\ref{fig:total}.
The open squares are the estimates $\lir^{\rm linear}$,
obtained from the linear relation [Eq.~(\ref{eq:12um})], normalized to $\lir$.
The vertical error bars correspond to the 95~\% confidence interval shown in 
Fig.~\ref{fig:12um}.
The linear relation gives reasonable values for normal galaxies.
}\label{fig:total_12um}
\end{figure}

\begin{figure}
\resizebox{\hsize}{!}{
\includegraphics[width=0.7\textwidth]{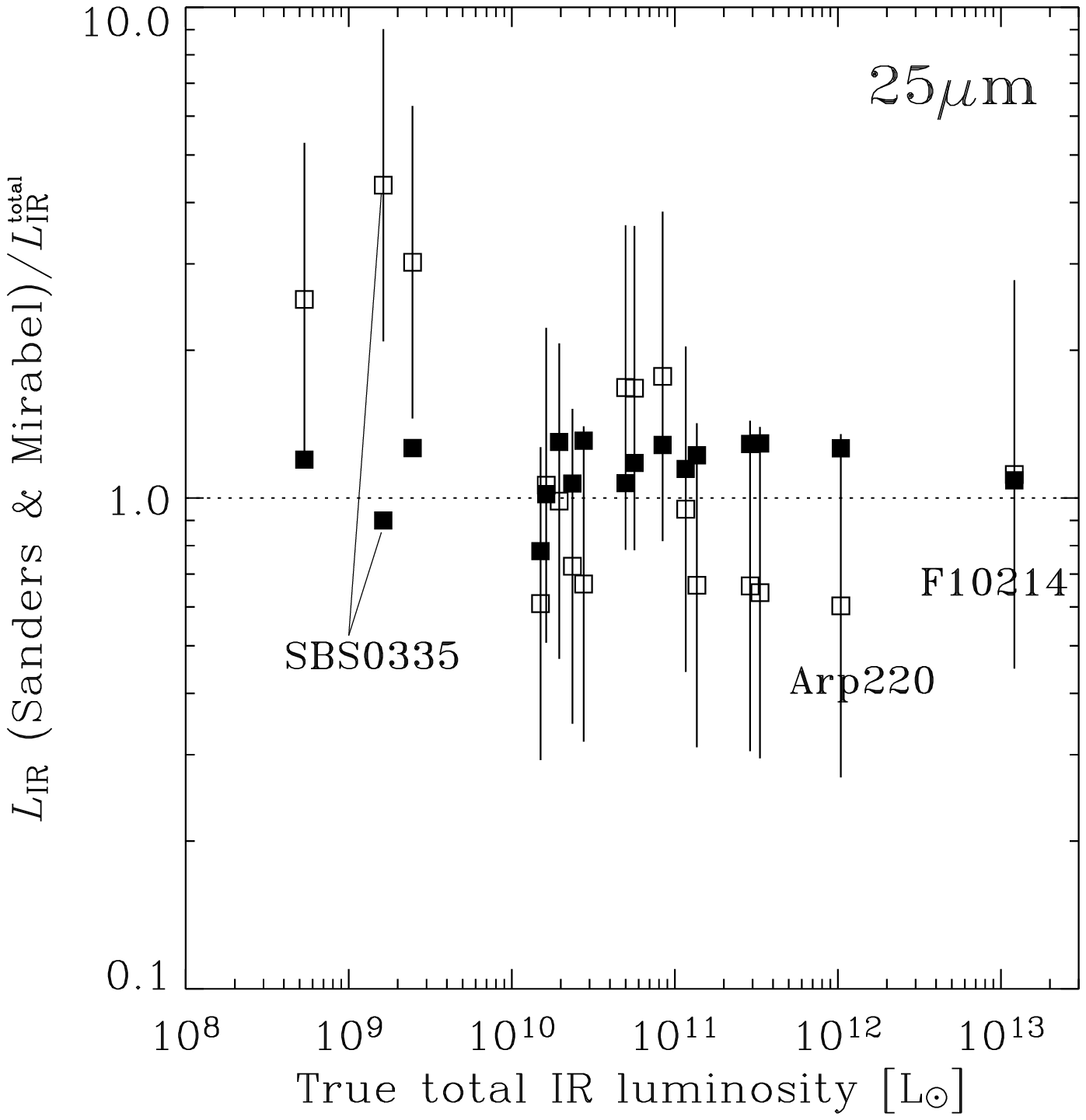}
}
\caption{The performance of the linear estimation formula obtained from the 
$L(25\,\mu\mbox{m})$--$\lir$ relation.
Symbols are as in Fig.~\ref{fig:total_12um}, except that
they are for $25\,\mu$m.
}\label{fig:total_25um}
\end{figure}

We plot the relation between the true integrated $\lirt$ 
and $\lir^{\rm linear}/\lirt$ in Figs.~\ref{fig:total_12um} 
and \ref{fig:total_25um}.
We also show the direct estimates from the formula of 
\citet{sanders96} using the four {\sl IRAS} flux densities (filled squares).
The ratios $\lir^{\rm linear}/\lirt$ are presented by
open squares with error bars that represent the 95~\% confidence interval.
In Fig.~\ref{fig:total_12um} the prediction is obtained from the 12-$\mu$m 
relation, while in Fig.~\ref{fig:total_25um} it is obtained from the 25-$\mu$m
relation.

In Fig.~\ref{fig:total_12um}, most of the normal galaxies give
reasonable agreement between $\lir$ and the estimates from the linear 
relation,  $\lir^{\rm linear}$.
However, the linear relation underestimates luminosities for 
three IR luminous galaxies (\object{NGC 2623}, \object{UGC 8387}, and \arp).
We also find that $\lir^{\rm linear}$ of \object{NGC 1569} is also 
smaller than the true value.
In fact, they have strongly extincted, red SEDs
(see Appendix~A), i.e., it is more IR-luminous than 
expected from their MIR luminosities.
For the other extreme, the linear relation gives acceptable estimates
(\sbs\ and \irasf) within the 95~\% confidence level.
Thus, we conclude that the 12-$\mu$m linear relation can be applicable for 
most of the variety of SEDs, except the extremely extinguished ones like \arp.
For such `red' galaxies, it gives a significant underestimation for $\lir$.

In Fig.~\ref{fig:total_25um}, in contrast, \arp\ and other red galaxies
are no longer serious outliers. 
On the other hand, \sbs\ significantly deviates upward from the true $\lir$.
Since \sbs\ has very hot dust emission \citep{dale01b}, 
the linear relation overestimates the $\lir$.
Anther two dwarf galaxies, \object{NGC 1569} and \iizw, are also
fairly overestimated because they also have warm dust emission.
However, the estimate for \irasf\ is excellent.
Hence, the linear relation between $L(25\,\mu\mbox{m})$ and $\lir$ tends to 
overestimate the $\lir$ for the galaxies with hot dust, but it works well 
for AGN-like SEDs, i.e., SEDs with a hot dust emission as well as with 
a FIR thermal emission.

\subsection{Formula for galaxies at $z \simeq 0.6$ based on 15-$\mu$m
luminosity}

Now we consider the higher-$z$ universe.
As mentioned in Sect.~\ref{sec:introduction}, our relations will be 
undoubtedly useful to estimate $\lir$ for galaxies detected in the
very deep {\sl Spitzer} MIPS data.

For galaxies at $z=1$, the $L(12\,\mu\mbox{m})$--$\lir$ linear relation 
itself can be used as an estimator of the total IR luminosity from 
the MIPS $24\,\mu$m band.
What should we do to estimate the total IR luminosity for galaxies at 
redshifts between 0 and 1?
In this subsection, we try to make a useful `interpolation' formula, 
which can be used to estimate the total IR luminosity for galaxies 
at $z \simeq 0.5\mbox{--}0.7$ in {\sl Spitzer} data.

The practical difficulty is the complexity of the MIR SED of galaxies.
At these wavelengths, we observe many aromatic band features
\citep[e.g.,][]{madden00}, thus, simple linear interpolation might not 
work well.
A more complex and continuous interpolation requires some kind of galaxy SED
model which is no longer free of assumptions, often not well-understood.
In this work, we stick to the empirical relationships directly obtained from 
observed datasets.
Thus, based on the {\sl ISO} deep 15-$\mu$m observations,
we try to find a relationship between $15\,\mu$m and $\lir$, 
since the observed wavelength of $24\,\mu$m corresponds to 
the emitted wavelength of $15\,\mu$m at $z=0.6$.
Although it cannot cover the whole range of $z=0\mbox{--}1$, it can be 
applied to a significant fraction of galaxies in this redshift range:
taking into account the photometric redshift uncertainty, we consider 
galaxies at $z\simeq 0.5\mbox{--}0.7$.
If we suppose a flux density limit of $200 \,\mu$Jy, the corresponding
luminosity at these redshifts will be $L(15\,\mu\mbox{m}) \simeq 
10^{9}\,L_\odot \simeq L_*$ \citep[see ][]{fang98}.
Hence, the fraction of galaxies at $z\simeq 0.5\mbox{--}0.7$ among
the detected galaxies will be 20--40~\%.

\subsubsection{Estimation formula for the total IR luminosity from 15-$\mu$m 
luminosity}\label{subsec:15um}

\citet{dale01a} provided average flux density ratios for {\sl IRAS} and
{\sl ISO} bands as a function of the ratio $S_{60}/S_{100}$.
It is well known that these flux density ratios depend on the $S_{60}/S_{100}$ 
ratio in general, so that the empirical SED models work well
\citep[e.g., ][]{dale01a,franceschini01,takeuchi01a,xu01,totani02,lagache03}.
For our purposes, however, the $S_{12}/S_{15}$ ratio only weakly depends on 
the $S_{60}/S_{100}$ ratio compared to other wavebands, because
the wavelength difference of these two bands is small.
We can also derive the formula for $15\,\mu$m from the 
$L(25\,\mu\mbox{m})$--$\lir$ relation
via $S_{15}/S_{25}$, however $S_{15}/S_{25}$ has a stronger and more systematic
dependence.
Since such a systematic dependence will result in a larger dispersion in 
the linear relation and reduce its reliability, we adopt $S_{12}/S_{15}$ 
for further discussion.

Then, considering the error of this ratio, we can safely use the average
value over the sample of \citet{dale01a} (their Table~1, column 8).
We found $\log \left(S_{12}/S_{15}\right) = 0.112$, which corresponds to 
$\log \left[L(12\,\mu\mbox{m})/L(15\,\mu\mbox{m})\right] = 0.209$.
Assuming that the slope of the MIR--total IR luminosity relation
does not change significantly between 12 and $15\,\mu$m, we obtain the 
following relation
\begin{eqnarray}
  \log \lir = 1.23 + 0.972 \log L(15\,\mu\mbox{m}) \;.\label{eq:mir_ir_15um}
\end{eqnarray}
The linear formula between $L(15\,\mu\mbox{m})$--$\lir$ luminosities 
[Eq.~(\ref{eq:mir_ir_15um})] shows a good agreement with the relation by 
direct fitting of the data proposed by \citet{chary01}, 
within the quoted error:
\begin{eqnarray}
  \log \lir = \left(1.05\pm0.174\right)+ 0.998 \log L(15\,\mu\mbox{m})\,.
\end{eqnarray}

\subsubsection{Examination of the $15\,\mu$m formula by observed galaxy sample}

In order to check the validity of Eq.~(\ref{eq:mir_ir_15um}), we use 
the quiescent galaxy sample in the Virgo cluster and the Coma/Abell 1367 
supercluster regions \citep{boselli03a}.
\citet{boselli04} have reported a good correlation between 
$L(15\,\mu\mbox{m})$ and $\lfir$ for the galaxies in the sample.
We again constructed a 'good quality' subsample with flux densities
in all the bands of {\sl IRAS} and {\sl ISO}.
We put a further constraint that the detected flux has quality flag 1
[$Q$ of \citet{boselli03a}: column~(14) in their Table~2] and examined
if the flux density suffers contamination by their close neighbors, 
and end up with a final subsample of 32 galaxies.

\begin{figure}
\resizebox{\hsize}{!}{
\includegraphics[width=0.7\textwidth]{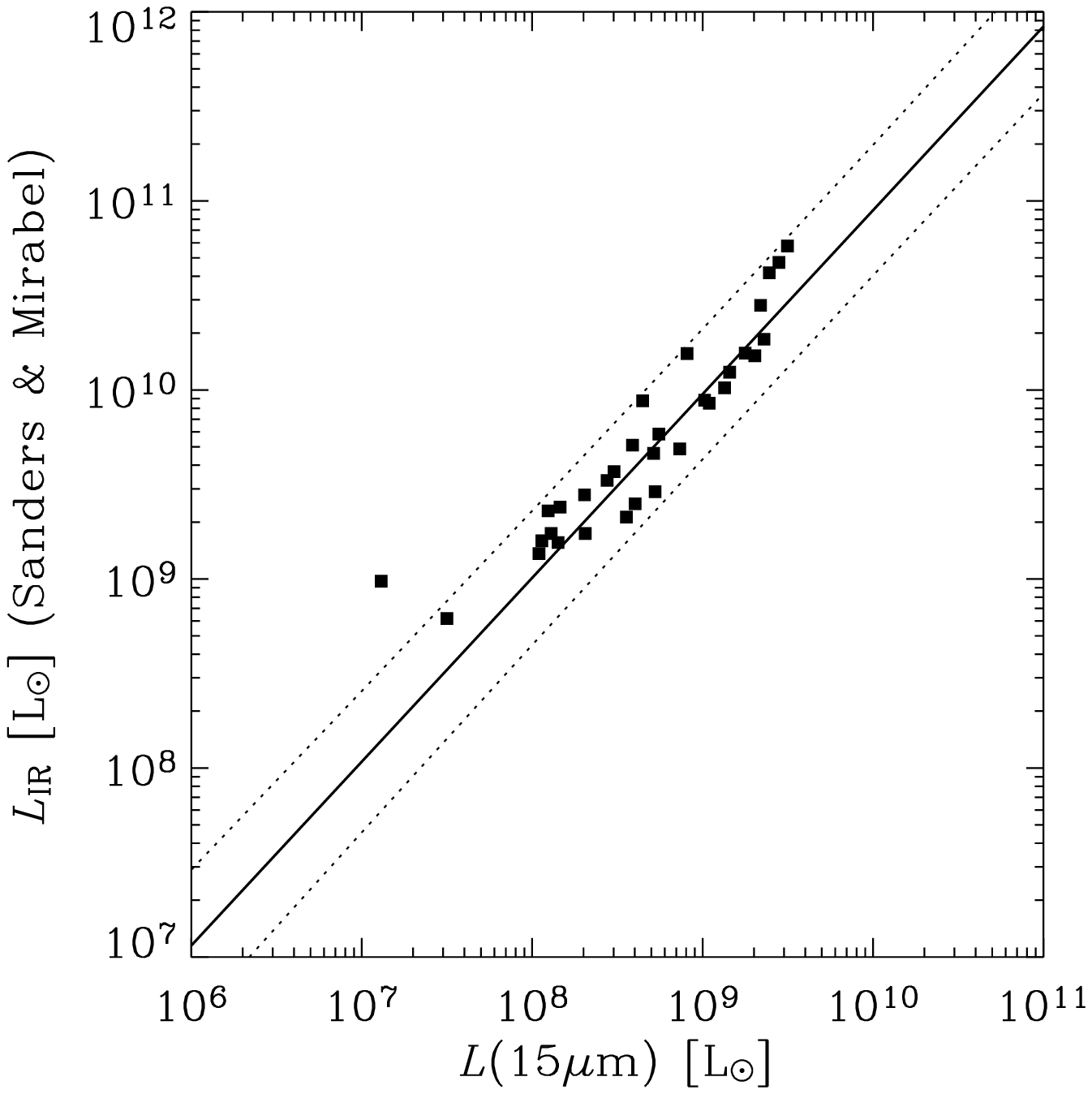}
}
\caption{The relation between $15\,\mu$m luminosity and $\lir$ for 
the quiescent galaxy sample of \citet{boselli03a} with our formula for
estimating $\lir$.
We also show the 95~\% confidence interval, which is the same as that obtained
for the $L(12\,\mu\mbox{m})$--$\lir$ relation.
}\label{fig:virgo_corrected}
\end{figure}

We plot this sample and our empirical formula (with 95~\% confidence
interval) in Fig.~\ref{fig:virgo_corrected}.
The formula is represented by the solid lines, and the confidence limits
are shown by dotted lines.
Indeed, 31 out of 32 galaxies lie in the confidence interval in each 
panel, i.e., the prediction from the formulae successfully work for 
$\sim 95\,\%$ of the sample.
Thus, we conclude that Eq.~\ref{eq:mir_ir_15um} is a reliable estimator of the
$\lir$ from 15-$\mu$m luminosity with an uncertainty of a factor of 4--5, 
and if the effect of the evolution is small, this relation can be used as 
an estimator of $\lir$ from the $24\,\mu$m luminosity of 
a galaxy at $z\simeq 0.6$.

However, we must keep in mind that there is clear evidence of a strong
evolution of galaxies \citep[e.g., ][]{takeuchi00,takeuchi01a,takeuchi03b} 
at $0<z<1$, and we expect a significant brightening of galaxies up to a
factor of a few at $z=0.5-0.6$ \citep[e.g.,][]{takeuchi01a,lagache03}.
Further investigation with physically-based models and high-$z$ observations
should be done in order to examine and/or modify the present formulae.

\section{Summary and conclusion}\label{sec:conclusion}

In this work, we first examined four IR luminosity estimators, 
$\lfir$ \citep{helou88}, 
$\ltir$ \citep{dale01a}, 
$\lrir$ \citep{dale02} and 
$\lir$ \citep{sanders96} with the observed SEDs of well-known galaxies.
We found that $\ltir$, $\lrir$, and $\lir$ correct the contribution from
the wavelengths missed by $\lfir$, but the latter two are better.
The estimator $\lir$ provides excellent estimates for a very wide variety of
galaxy SEDs, from SEDs indicative of very hot dust
(e.g., \sbs\ and \irasf) to very extinguished SEDs and/or cold dust
emission (e.g., \arp).
We also note that the performance of $\lrir$ is almost as good as that 
of $\lir$.

Using $\lir$, we then statistically analyzed the {\sl IRAS} PSC$z$ galaxy 
sample \citep{saunders00} and found useful formulae relating the MIR 
monochromatic luminosities [$L(12\,\mu\mbox{m})$ and $L(25\,\mu\mbox{m})$], 
and $\lir$.
For this purpose we constructed a subsample of 1420 galaxies with all 
four {\sl IRAS} band (12, 25, 60, and $100\,\mu$m) flux densities.
We found linear relations between $\lir$ and MIR luminosities, 
$L(12\,\mu\mbox{m})$ and $L(25\,\mu\mbox{m})$.
The prediction error with 95~\% confidence level is a factor of 4--5.
Hence, these formulae are useful for the estimation of the total
IR luminosity $\lirt$ only from $12\,\mu$m or $25\,\mu$m observations.

We further tried to make an `interpolation' formula for galaxies 
in the middle of $z=0$ and 1.
For this purpose we construct the formula of the relation between 15-$\mu$m
luminosity and the total IR luminosity using the flux density ratio of
\citet{dale01a}.
The obtained formula well reproduced the observed relation in the
sample of \citet{boselli03a}.
We conclude that the 15-$\mu$m formula can be used as an estimator of the
total IR luminosity from $24\,\mu$m observations of galaxies at $z \simeq 0.6$.

\appendix

\section{SED of our well-known galaxy sample}\label{sec:sed}

\begin{figure}
\resizebox{\hsize}{!}{
\includegraphics[width=0.7\textwidth]{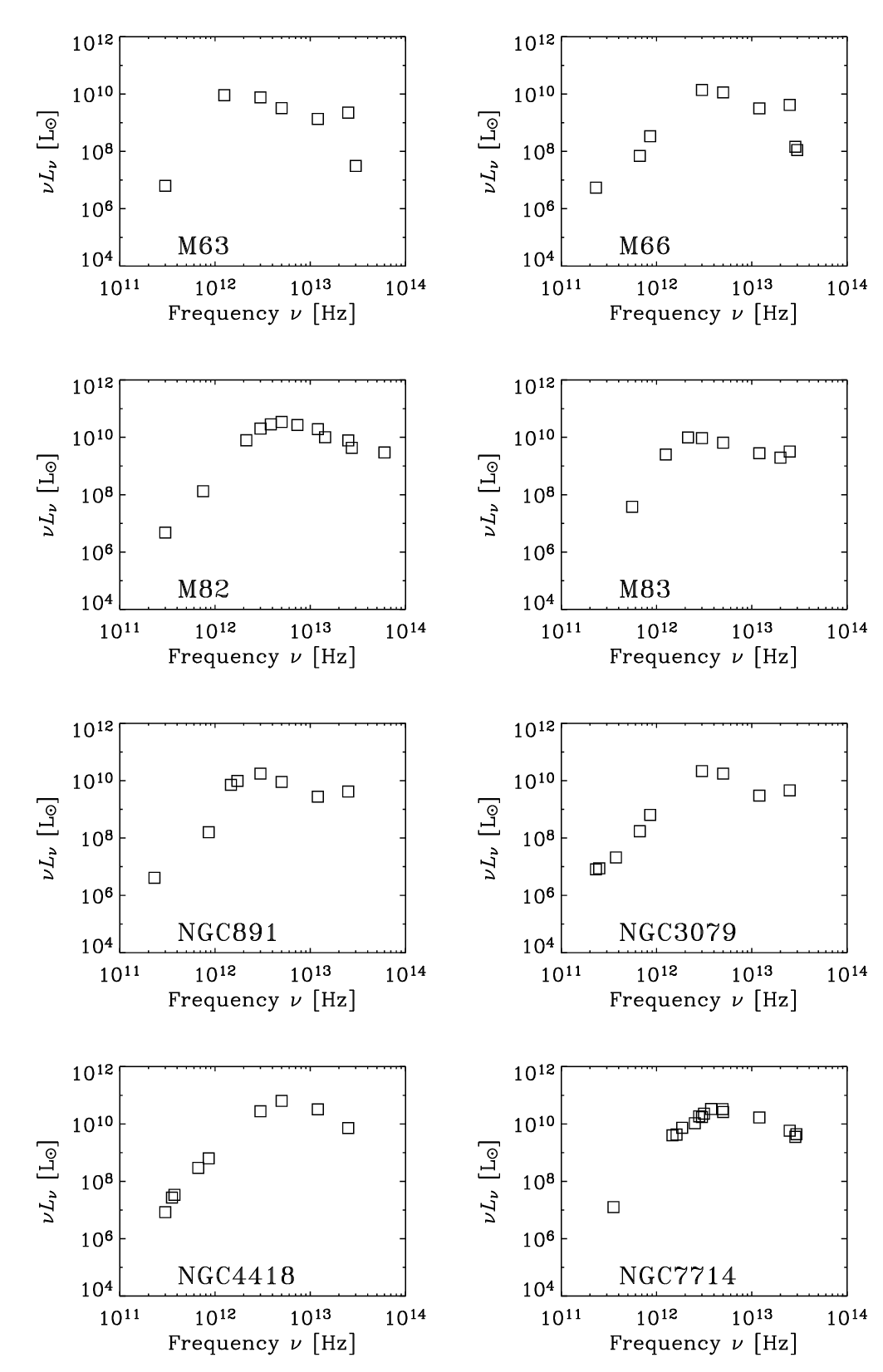}
}
\caption{The galaxy sample with total IR luminosities, $\lirt=
10^{10}\,L_\odot \mbox{--} 10^{11}\,L_\odot$. 
}\label{fig:sed_normal}
\end{figure}

\begin{figure}
\resizebox{\hsize}{!}{
\includegraphics[width=0.7\textwidth]{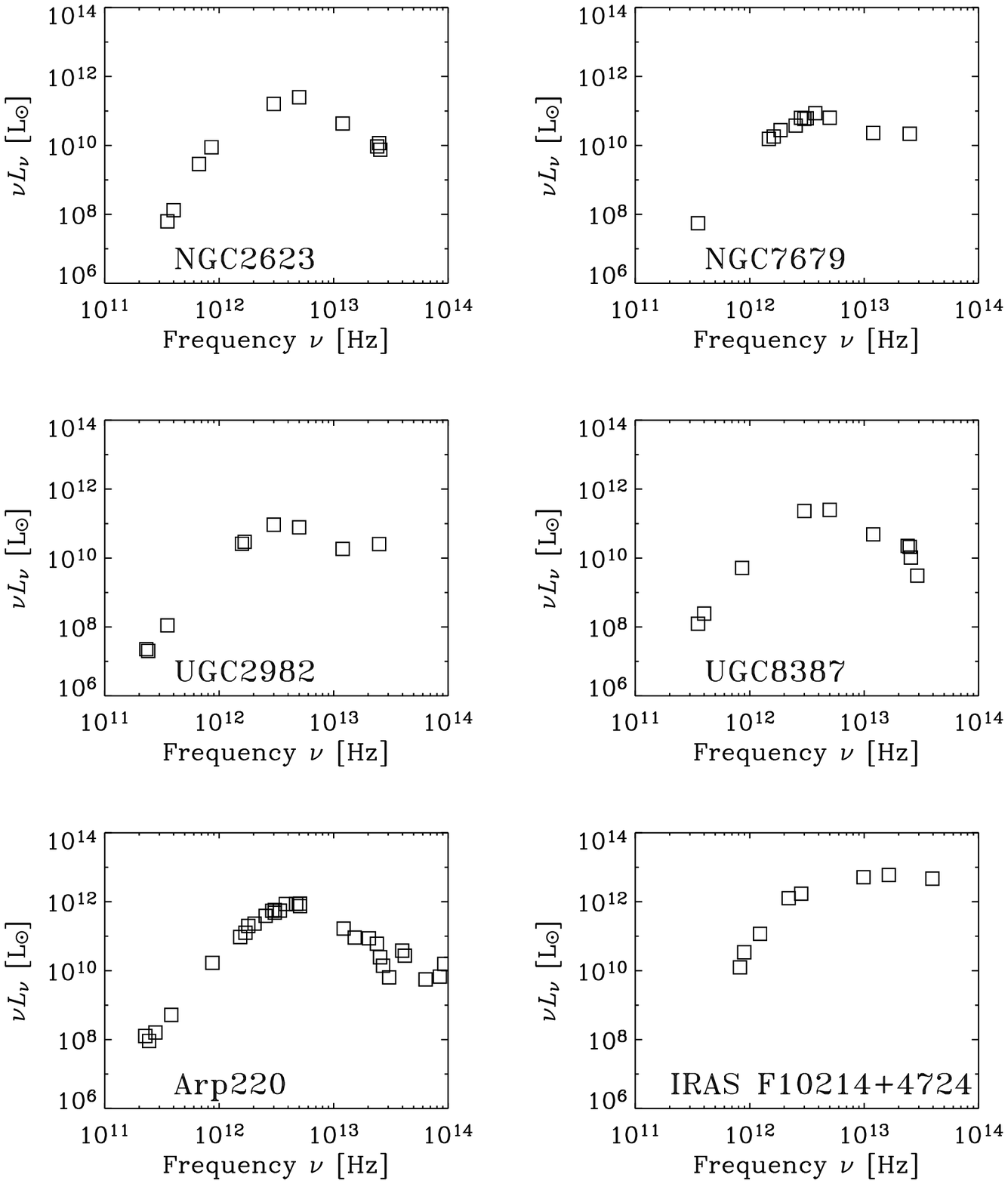}
}
\caption{The IR-luminous galaxy sample, with $\lirt > 10^{11}\,L_\odot$. 
}\label{fig:sed_lirg}
\end{figure}

\begin{figure}
\resizebox{\hsize}{!}{
\includegraphics[width=0.7\textwidth]{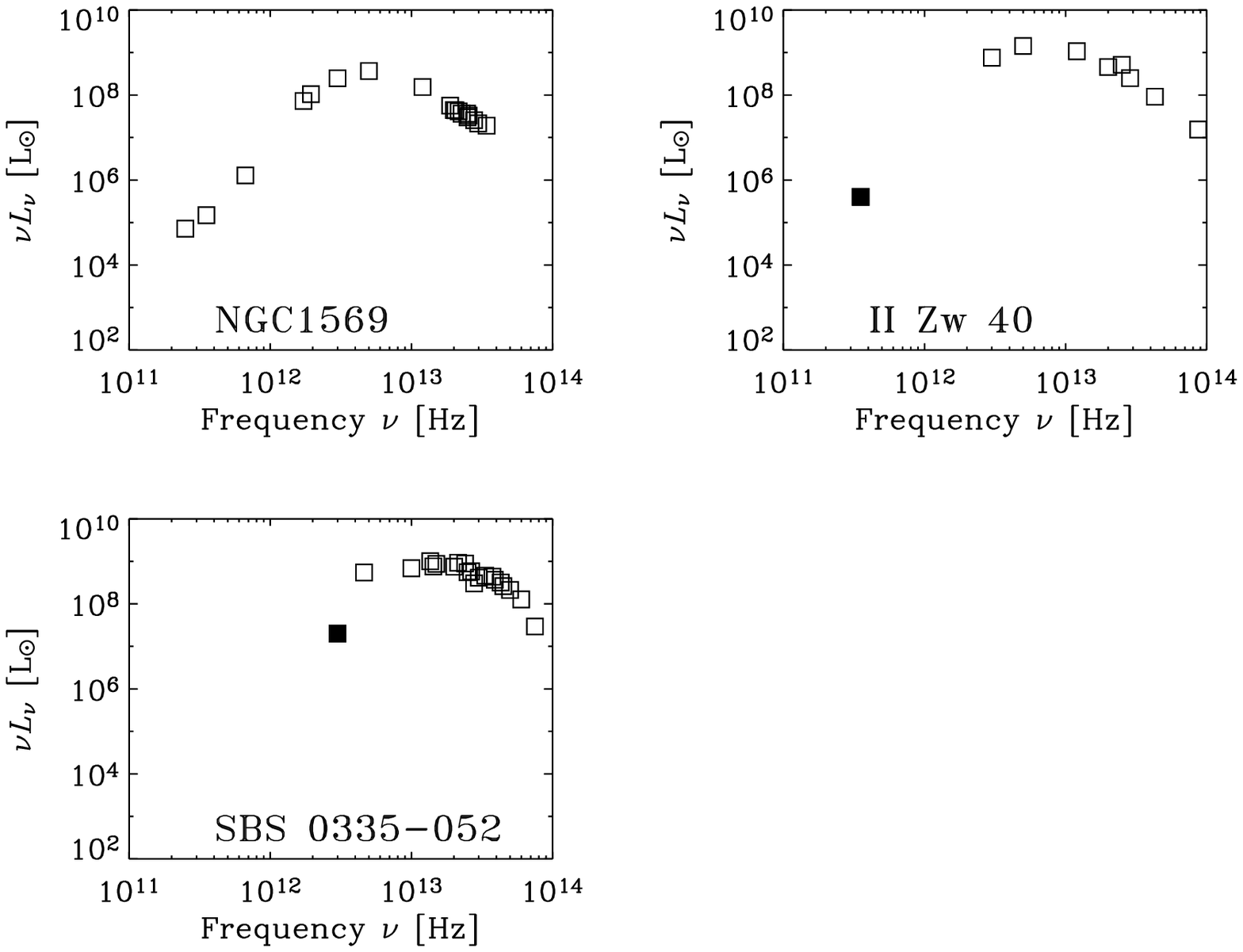}
}
\caption{The dwarf galaxy sample, with $\lirt < 10^{10}\,L_\odot$. 
No observational data are available at FIR--submm for \iizw\ and \sbs, 
we extrapolated the SEDs by the SED model for a young galaxy presented
by Takeuchi et al.~(2003).
The extrapolated data are represented by filled squares.
}\label{fig:sed_dwarf}
\end{figure}

In Appendix~\ref{sec:sed}, we present all the observed SEDs of galaxies
we used in examining the performance of the total IR luminosity estimators.
We show the normal galaxy sample with $10^{10}\,L_\odot < \lirt < 
10^{11}\,L_\odot$ in Fig.~\ref{fig:sed_normal}, IR-luminous sample in 
Fig.~\ref{fig:sed_lirg}, and dwarf sample in Fig.~\ref{fig:sed_dwarf}.
Among the dwarf sample, for \sbs\ and \iizw, the interpolated points are
represented by filled squares (see main text).

\section{Regression analysis}\label{sec:regression}

We made a regression analysis for the logarithms of $L(\lambda)$.
It should be noted here that we are interested in estimating the total 
IR luminosity $\lirt$ from the MIR luminosity.
Then, in the regression analysis, the uncertainty that we need is
the so-called {\sl prediction error}, not the error of the regression
parameters.
We represent the linear regression model as
\begin{eqnarray}
  y=\beta_0+\beta_1 x + \varepsilon \;,
\end{eqnarray}
where 
\begin{eqnarray}
  \varepsilon \sim N(0,\sigma^2)\;.
\end{eqnarray}
Here the symbol `$\sim$' means that the stochastic variable $\varepsilon$
obeys a Gaussian distribution $N$ with a mean 0 and dispersion $\sigma^2$.
The following estimators $b_0$ and $b_1$ are known as the 
best unbiased estimators\footnote{That is,  
$\expc{b_0} = \beta_0$ and $\expc{b_1}=\beta_1$, where $\expc{x}$
represents the expectation value of a stochastic variable $x$, and the variance
is the smallest among the estimators.} for $\beta_0$ and $\beta_1$:
\begin{eqnarray}
  b_1 \equiv \frac{\sum_{i=1}^n (x_i-\bar{x})(y_i-\bar{y})}{
    \sum_{i=1}^{n}(x_i-\bar{x})^2} 
\end{eqnarray}
and
\begin{eqnarray}
  b_0 \equiv \bar{y} - b_1 \bar{x} \;,
\end{eqnarray}
where $\bar{x}$ and $\bar{y}$ are the sample mean of $x_i$ and $y_i$, 
respectively.
The dispersion of $b_0$ and $b_1$ shows the statistical uncertainty of 
{\sl parameters}.
However, in a practical application, we need a dispersion of the estimation
value $\hat{y}(x_0)$ (here hat means that the value is the 
predicted one and not the sample value which would be obtained in 
a potential new observation at $x_0$)
for a certain value $x_0$ of the independent variable 
$x$, in the sense that if we could repeat an observation $n$ times,
we want an interval within which, for example,
95~\% of the prediction values $\hat{y}(x_0)$ lie.
This range is the {\sl prediction error}, and can be evaluated by the formula
\begin{eqnarray}\label{eq:prediction_error}
  V_{\rm p} &\equiv& \var{y(x_0)-\hat{y}(x_0)} \nonumber \\
  &=& \var{y(x_0)} + \var{\hat{y}(x_0)} \nonumber \\
  &=& \left[ 1+\frac{1}{n}+
    \frac{(x_0-\bar{x})^2}{\sum_{i=1}^n(x_i-\bar{x})^2}\right]\sigma^2 \;,
\end{eqnarray}
where the symbol $\var{x}$ signifies the variance of a stochastic 
variable $x$.
The second line of eq.~(\ref{eq:prediction_error}) follows from the 
statistical independence between $y(x_0)$ and $\hat{y}(x_0)$.
Observationally, we should replace $\sigma^2$ with its unbiased estimator,
$V_{\rm e}$ [whose unit is a square of dex (an order of magnitude
in luminosity)], obtained as
\begin{eqnarray}
  V_{\rm e} = \frac{\sum_{i=1}^n (y_i - \hat{y}_i)^2}{n-2} \;.
\end{eqnarray}
The 95~\% confidence interval for the regression line is, then, represented by 
$\hat{y}(x_0) \pm 2.228 \sqrt{V_{\rm p}}$.
For further statistical discussions, see e.g., Stuart, Ord, \& Arnold (1999).

\begin{acknowledgements}
We offer our thanks to Daniel Dale, the referee, for his useful
comments that much improved the clarity of this paper.
We also thank Akio K.\ Inoue, Akihiko Ibukiyama, and Luca Cortese for their 
helpful comments and suggestions.
This research has made use of the NASA/IPAC Extragalactic
Database (NED) which is operated by the Jet Propulsion Laboratory, Caltech,
under contract with the National Aeronautics and Space Administration.
We made extensive use of the NASA Astrophysics Data System.
TTT has been supported by the Japan Society for the Promotion of Science.
\end{acknowledgements}

\end{document}